\documentclass[letterpaper, 10 pt, conference]{ieeeconf}  

\IEEEoverridecommandlockouts                              
\makeatletter
\let\NAT@parse\undefined
\makeatother

\usepackage[dvipsnames]{xcolor}

\newcommand*\linkcolours{ForestGreen}

\usepackage{times}
\usepackage{graphicx}
\usepackage{amssymb}
\usepackage{gensymb}
\usepackage{amsmath}
\usepackage{float}
\usepackage{breakurl}

\usepackage{url,hyperref}
\hypersetup{
colorlinks,
linkcolor=\linkcolours,
citecolor=\linkcolours,
filecolor=\linkcolours,
urlcolor=\linkcolours}

\usepackage{algorithm}
\usepackage{algorithmic}

\usepackage[labelfont={bf},font=small]{caption}
\usepackage[none]{hyphenat}

\usepackage{mathtools, cuted}

\usepackage[noadjust, nobreak]{cite}

\usepackage{tabularx}
\usepackage{amsmath}

\usepackage{float}

\usepackage{pifont}

\newcolumntype{Y}{>{\centering\arraybackslash}X}

\usepackage[]{placeins}

\usepackage{placeins}

\usepackage{tikz}

\usepackage[framemethod=tikz]{mdframed}

\usepackage{afterpage}

\usepackage{stfloats}

\usepackage{atbegshi}
\newcommand{\handlethispage}{}
\newcommand{\discardpagesfromhere}{\let\handlethispage\AtBeginShipoutDiscard}
\newcommand{\keeppagesfromhere}{\let\handlethispage\relax}
\AtBeginShipout{\handlethispage}

\usepackage{comment}

\title{\LARGE \bf
Laser Cooling of Silica Glass
}

\author{Esmaeil Mobini$^{1,2}$, Saeid Rostami$^{1}$,  Mostafa Peysokhan$^{1,2}$, Alexander Albrecht$^{1}$, \\
Stefan Kuhn$^{3}$, Sigrun Hein$^{3}$, Christian Hupel$^{3}$, Johannes Nold$^{3}$, Nicoletta Haarlammert$^{3}$,\\ 
Thomas Schreiber$^{3}$, Ramona Eberhardt$^{3}$, Andreas T\"unnermann$^{3,4}$,\\ 
Mansoor Sheik-Bahae$^{1}$, and Arash Mafi$^{1,2,*}$
\thanks{$^{1}$Department of Physics \& Astronomy, University of New Mexico, Albuquerque, New Mexico 87131, USA}%
\thanks{$^{2}$Center for High Technology Materials, University of New Mexico, Albuquerque, New Mexico 87106, USA}
\thanks{$^{3}$Fraunhofer Institute for Applied Optics and Precision Engineering, Albert-Einstein-Str. 7, 07745 Jena, Germany}
\thanks{$^{4}$Institute of Applied Physics, Abbe Center of Photonics, Friedrich-Schiller-Universit\"at, Albert-Einstein-Str. 15, 07745 Jena, Germany}
\thanks{$^{*}$Email: mafi@unm.edu}%
}

\begin{document}

\maketitle
\thispagestyle{empty}
\pagestyle{empty}

\begin{abstract}
Laser cooling of a solid is achieved when a coherent laser illuminates the material in the red tail of its absorption spectrum, 
and the heat is carried out by anti-Stokes fluorescence of the blue-shifted photons. 
Solid-state laser cooling has been successfully demonstrated in several materials, including rare-earth-doped crystals and glasses. 
Silica glass, being the most widely used optical material, has so far evaded all laser cooling attempts. 
In addition to its fundamental importance, many potential applications can be conceived
for anti-Stokes fluorescence cooling of silica. These potential applications range from the substrate cooling of optical circuits 
for quantum information processing and cryogenic cooling of mirrors in high-sensitivity interferometers for gravitational wave detection 
to the heating reduction in high-power fiber lasers and amplifiers. Here we report the net cooling of high-purity Yb-doped silica glass samples 
that are primarily developed for high-power fiber laser applications, where special care has been taken in the fabrication process to reduce 
their impurities and lower their parasitic background loss. The non-radiative decay rate of the excited state in Yb ions is very small in 
these glasses due to the low level of impurities, resulting in near-unity quantum efficiency. 
We report the measurement of the cooling efficiency as a function of the laser wavelength, from which the quantum efficiency of the silica glass is calculated.
\end{abstract}

In solid-state laser cooling, anti-Stokes fluorescence removes heat from the material, resulting in net refrigeration. 
Pringsheim first proposed it in 1929~\cite{Pringsheim1929} and Epstein et al. reported its first experimental confirmation in Yb-doped ZBLAN in 1995~\cite{epstein1995observation}. 
Multiple experiments have since confirmed solid-state laser cooling; they have focused on three broad classes of solids: 
crystals, semiconductors, and glasses. 
Laser cooling of crystals has been the most successful so far~\cite{seletskiy2010laser,nemova2010laser,seletskiy2016laser}; the record cooling to 91\,K of a 10\,mol\% Yb:YLF crystal was reported at the 
University of New Mexico in 2016~\cite{melgaard2016solid}. The only reported laser cooling of semiconductors is that of a CdS nanobelt in 2013 by 40\,K~\cite{zhang2013laser}, 
but the validity of their results has been questioned recently~\cite{morozov2019can}. Several glasses have been successfully 
cooled~\cite{gosnell1999laser,fernandez2000,PhysRevLett.85.3600,thiede2005cooling,fernandez2006anti,nguyen2013towards,peysokhan2019laser}
since the first experimental report by Epstein et al.~\cite{epstein1995observation}. 
However, attempts to cool silica glass, which is arguably the 
most versatile optical material, have so far been unsuccessful~\cite{PhysRevApplied-E.Mobini,JennyKnall-SPIE}. Here, we report the first laser cooling of Yb-doped silica glass.

The perennial failure in the laser cooling of silica glass led some to even question its possibility; the main skepticism focused on whether it would be possible for the 
Yb-doped silica glass to have a sufficiently small non-radiative decay rate of the Yb excited-state population to achieve a near-unity internal quantum efficiency. 
This was examined recently in a spectroscopic study of the Yb-doped silica glass and by looking into the potential decay channels of the Yb excited-state population; 
it was concluded that there is no a priori reason to reject the possibility of laser cooling for the high-purity Yb-doped silica glass~\cite{PhysRevApplied-E.Mobini}. 
However, it was predicted that for an improved laser cooling, the glass host must be codoped with modifiers such as Al, to mitigate the
quenching-induced non-radiative decay~\cite{laegsgaard2002dissolution,arai1986aluminum}.

Advancements in solid-state laser cooling may eventually lead to all-optical compact and vibration-free cryocoolers that can reduce the thermal noise in 
semiconductor-based single-photon detectors or quantum information processing circuits~\cite{seletskiy2016laser}. 
Another important application is for radiation-balanced fiber lasers (RBFLs), where the cooling from anti-Stokes fluorescence offsets the waste heat generation 
in the laser~\cite{bowman1999lasers,nemova2009athermal,bowman2010minimizing,nemova2011radiation,Esmaeil2018josabRBL,Yang:19}. 
Rare-earth-doped crystals like Yb:YLF have proven to be the best materials of 
choice for laser cooling because they have a small inhomogeneous broadening of the absorption lines and a high ion solubility that leads to a higher cooling efficiency~\cite{seletskiy2016laser,seletskiy2010laser}. 
However, the incompatibility of doped crystals with silicon-based devices may limit their potential applications~\cite{seletskiy2016laser}. 
ZBLAN glass is another successful cooling-grade material, but its low mechanical and chemical stability limits its application for silicon photonics or RBFLs.
On the other hand, Yb-doped silica glass is the material of choice for high-power fiber lasers and is commonly used as the substrate in 
silicon photonics~\cite{zhu2010high,Mobini:19,jalali2006silicon,soref2010mid}. 
Therefore, potential applications, especially for RBFLs in the near-term and photonic-device cooling in the long-run, are strong motivations for the laser cooling of silica 
glass beside the scientific curiosity.   

\begin{figure*}[t]
\centering
\includegraphics[width=0.8\textwidth]{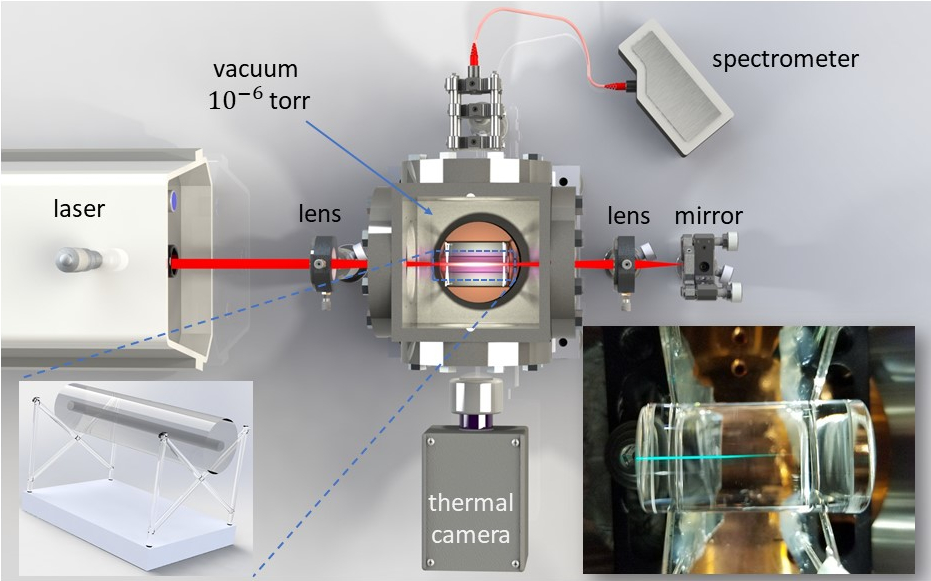}
\caption{{\bf Schematic of the LITMoS test setup.} A wavelength-tunable CW Ti:Sapphire laser is coupled by a lens of focal length f=10\,cm into the 
silica glass preform through a side window mounted on the vacuum chamber. 
The transmitted light gets reflected back by a highly-reflective mirror and is coupled back into the preform again by another lens of focal length f=10\,cm. 
The lower left inset shows a sketch of the 
Yb-doped silica glass preform supported by a set of thin silica fibers to minimize the heat load, while the lower right inset shows the actual image of preform supported on the thin silica fibers.}
\label{Fig:setup}
\end{figure*}

The primary focus of this Letter is to investigate the laser cooling of silica glass. In particular, we will determine the wavelength dependence of the cooling efficiency 
of our Yb-doped silica glass samples as a function of the pump laser wavelength to observe their transition from the heating to cooling regime.
The cooling efficiency, $\eta_c$, is defined as the net power density (per unit volume) extracted from the material ($p_{net}$) per unit power density absorbed or scattered
($p_{abs}$): $\eta_c=p_{net}/p_{abs}$. The cooling efficiency can be expressed as~\cite{NP6932007,seletskiy2016laser} (see Supplementary Information)
\begin{align}
\label{eq:etac}
\eta_c(\lambda_p)=\dfrac{\lambda_p}{\lambda_f}\,\eta_{ext}\,\eta_{abs}-1,
\end{align}
where  $\lambda_{f}$ is the mean fluorescence wavelength and $\lambda_{p}$ is the laser pump wavelength. 
$\eta_{ext}$ is the external quantum efficiency and $\eta_{abs}$ is the absorption efficiency; they are defined as: 
\begin{align}
\label{Eq:etaq}
&\eta_{ext}=\frac{\eta_{e}W_{r}}{W_{tot}},\quad W_{tot}=\eta_{e}W_{r}+W_{nr},\\
\label{Eq:etaabs}
&\eta_{abs}(\lambda_{p})=\frac{\alpha_{r}(\lambda_{p})}{\alpha_{r}(\lambda_{p})+\alpha_{b}},
\end{align}
where $W_{r}$, $W_{nr}$, and $W_{tot}$ are radiative, non-radiative, and total decay rates of the excited state, respectively, 
and $\eta_{e}$ is the fluorescence extraction efficiency. $\alpha_{b}$ is the background absorption coefficient, and 
$\alpha_{r}$ is the resonant absorption coefficient.
In practice, both $\eta_{ext}$ and $\eta_{abs}$ must be very close to unity to 
observe laser cooling, because $\lambda_p$ cannot be much longer than $\lambda_f$ to keep $\alpha_{r}(\lambda_p)$ sufficiently large for a
near unity value of $\eta_{abs}$.

It was recently shown in Ref.~\cite{PhysRevApplied-E.Mobini} that it is possible for the Yb excited-state population
to have a small non-radiative decay rate in a silica glass host, i.e. $W_{nr}\ll W_r$. Therefore, the external quantum efficiency 
can be near unity as long as $\eta_{e}\approx 1$. To revisit the arguments presented in Ref.~\cite{PhysRevApplied-E.Mobini}, note that
the non-radiative decay rate, $W_{nr}$, can be divided into two separate parts: the multiphonon decay rate ($W_{mp}$) and 
the sum of other non-radiative decay rates ($W_i$) for those channels that are related to the 
concentration quenching effect, i.e., $W_{nr}=W_{mp}+\Sigma_i W_{i}$~\cite{auzel2003radiation,PhysRevApplied-E.Mobini}. 
Using the energy-gap law, we showed that the multiphonon decay rate of silica glass is $W_{mp}^{\rm silica}\approx 10^{-8}\,{\rm s}^{-1}$, while
that of ZBLAN is $W_{mp}^{\rm ZBLAN}\approx 10^{-4}\,{\rm s}^{-1}$; therefore, as far as the multiphoton non-radiative decay rate is concerned, 
Yb-doped silica glass is a better material than ZBLAN for optical refrigeration~\cite{PhysRevApplied-E.Mobini}.

The non-radiative decay channels related to the concentration quenching are mainly due to the dipole-dipole interactions between Yb ions and impurities, which include
${\rm OH^{-}}$, transition metals, and undesirable rare-earth ions; as well as Yb-Yb interactions in Yb ion clusters. 
Developing a high-purity Yb-doped silica glass is therefore required to avoid the interactions between the Yb ions and impurities~\cite{PhysRevApplied-E.Mobini}.
Additionally, to ensure that Yb ion clustering is suppressed and to further mitigate Yb-impurity interactions, it is imperative for the 
Yb ion density to remain below the critical ion concentration~\cite{auzel2003radiation}. It is known that the ion solubility of the silica glass is quite 
low, i.e., for pure silica glass the critical quenching concentration is $N_{c}\approx 10^{25}$\,m$^{-3}$ or lower~\cite{barua2008influences}. 
However, by using modifiers such as Al and P, the quenching concentration of silica glass can be increased by an order of magnitude~\cite{laegsgaard2002dissolution,arai1986aluminum}. To prevent
concentration quenching and achieve $\eta_{ext}\approx 1$, it is necessary to keep the Yb ion density below $N_c$. Quite possibly,
this issue has been one of the main reasons behind the previously failed attempts in laser cooling of the Yb-doped silica glass~\cite{JennyKnall-SPIE}.
The Yb-doped silica glass samples that are studied in this Letter are all high-purity and are doped with modifiers to increase the Yb ion solubility~\cite{S.Kuhn-SPIE-2019}. The 
parasitic background absorption ($\alpha_b$) in these glasses is sufficiently low to ensure that $\eta_{abs}\approx 1$, as is required to achieve laser cooling.

For the laser cooling experiments, we used three different samples of Yb-doped silica glass optical fiber preforms (see Methods). 
We refer to these preforms as sample A, sample B, and sample C, respectively.
These preforms are Yb-doped only in the core and their characteristics are listed in Table.~\ref{Tab:PreformsFeat}.
\begin{table}[htp]
   \caption{Characteristics of the Yb-doped silica glass preforms.}
\begin{center}
\label{Tab:PreformsFeat}
 \renewcommand{\arraystretch}{1.3}
\begin{tabular}{ | p{4pt} | p{24pt} | p{14pt} | p{14pt} | p{14pt} | p{14pt} | p{14pt} | p{14pt} | p{14pt} |}
      \hline
      \rotatebox[origin=c]{90}{\parbox[c]{1.65cm}{\centering sample}} & 
      \rotatebox[origin=c]{90}{\parbox[c]{1.65cm}{\centering codopants}} &
      \rotatebox[origin=c]{90}{\parbox[c]{1.65cm}{\centering Yb$_2$O$_3$ [mol\%]}} & 
      \rotatebox[origin=c]{90}{\parbox[c]{1.65cm}{\centering Yb density [$10^{25}{\rm m}^{-3}$]}} &
      \rotatebox[origin=c]{90}{\parbox[c]{1.65cm}{\centering OH conc. [ppm]}} & 
      \rotatebox[origin=c]{90}{\parbox[c]{1.65cm}{\centering core diam. [mm]}} & 
      \rotatebox[origin=c]{90}{\parbox[c]{1.65cm}{\centering clad diam. [mm]}} & 
      \rotatebox[origin=c]{90}{\parbox[c]{1.65cm}{\centering length [mm]}} & 
      \rotatebox[origin=c]{90}{\parbox[c]{1.65cm}{\centering $\alpha_{b}$(1200\,nm) [dB/km]}} \\ 
      \hline
       A & Al,~P     & 0.12  & 5.3 & 3.0 & 1.7 & 10.7 & 28.6 & 10 \\
      \hline
       B & Al,~F     & 0.10  & 4.4 & 1.5 & 2.6 & 13.8 & 28.7 & 9 \\
      \hline
       C & Al,~F,~Ce & 0.13  & 5.7 & 1.5 & 3.1 & 14.8 & 27.8 & 5 \\ \hline
\end{tabular}
\end{center}
\end{table}

To investigate laser cooling and obtain the cooling efficiency, $\eta_{c}$, of the Yb-doped silica glass preforms as a function of the laser pump wavelength, 
we perform the Laser-Induced Thermal Modulation Spectroscopy (LITMoS) test on all three samples~\cite{seletskiy2016laser,Rostami:19} (see Supplementary Information). 
The LITMoS test setup is shown in Fig.~\ref{Fig:setup}.
The samples are held by a set of silica fibers inside a vacuum chamber with the pressure of $10^{-6}$\,Torr to eliminate the conductive and convective 
heat-loads on the samples, so the black body radiation remains the only source of heating from the environment. 
The samples are pumped by a wavelength-tunable continuous wave (CW) Ti-Sapphire laser ($980\,{\rm nm}<\lambda_p<1070\,{\rm nm}$) and the laser light 
passes through each sample twice using an external mirror. The thermal images and spectral features of the samples are captured through a set of 
thermally transparent KCl salt windows mounted in the chamber.
The changes of the temperatures are recorded by a thermal camera. To calculate the mean fluorescence wavelength, the samples are initially pumped at $\lambda_p$=1030\,nm. 
The fluorescence emission then is captured with an optical spectrum analyzer. The calculated mean fluorescence wavelengths of the samples A, B and C are found to be 
$\lambda_{f}^{\rm A}$=1010\,nm, $\lambda_{f}^{\rm B}$=1008\,nm, and $\lambda_{f}^{\rm C}$=1008\,nm, respectively~\cite{PhysRevApplied-E.Mobini} (see Supplementary Information).


\begin{figure}[ht]
\centering
\includegraphics[width=0.95\columnwidth]{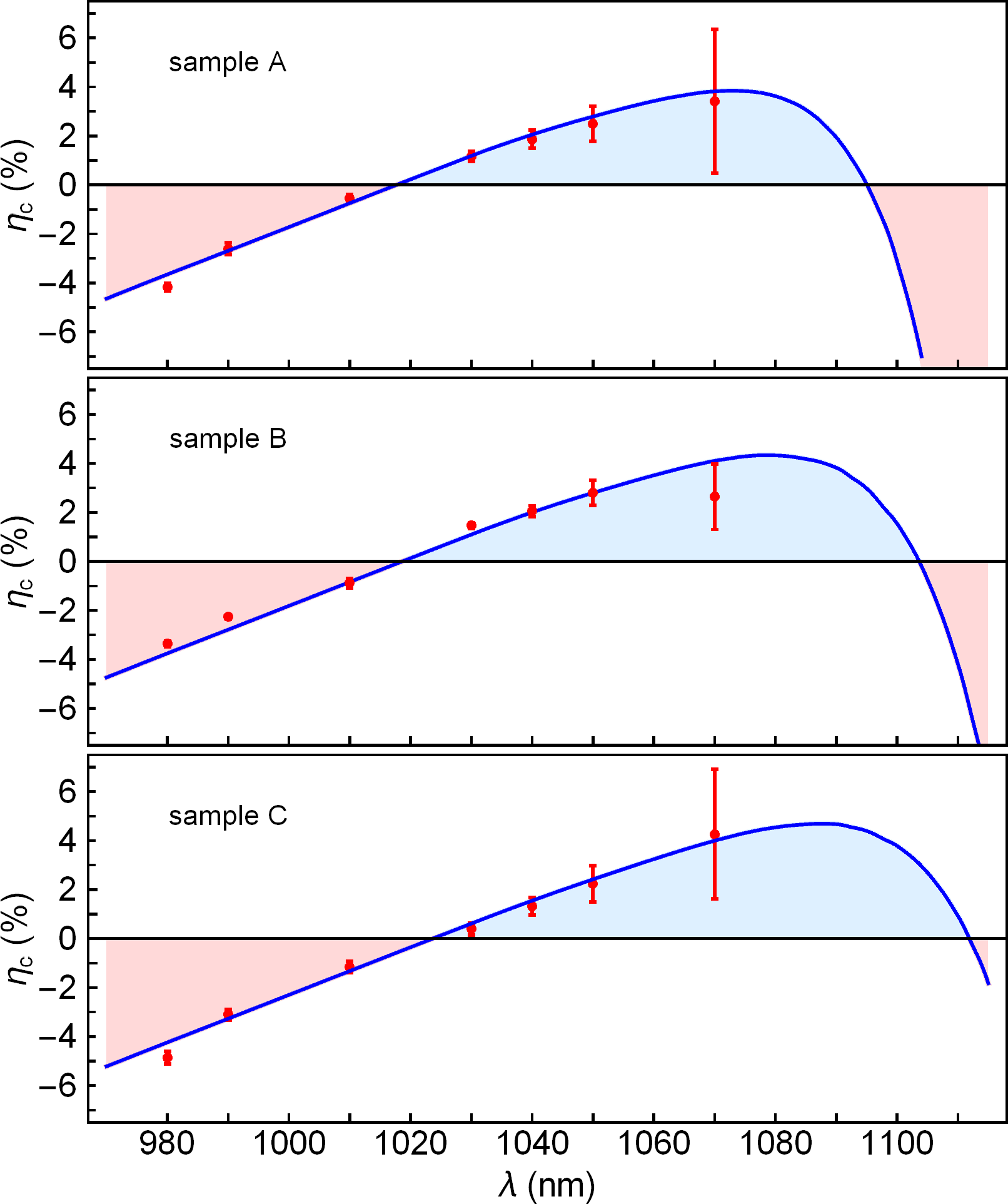}
\caption{{\bf Measurement of the cooling efficiency.}
In each subfigure, the red dots with error-bars represent the measured values of the cooling efficiency for samples A, B, and C, respectively.
The blue curved line is a fitting of Eq.~\ref{eq:etac} to the experimental measurements. 
}
\label{Fig:cooleff-All}
\end{figure}

Figure~\ref{Fig:cooleff-All} shows the obtained cooling efficiencies of the three samples obtained from the LITMoS test. In each subfigure corresponding to the particular sample A, B, or C, 
the pump laser wavelength is gradually increased; once it becomes longer than the mean fluorescence wavelength, the anti-Stokes fluorescence begins to extract heat 
from the sample until the cooling efficiency becomes positive, indicating the net laser cooling. As can be seen in Fig.~\ref{Fig:cooleff-All},
all three samples have been laser cooled. By fitting Eq.~\ref{eq:etac} to the experimental results and using the values of $\alpha_b$ reported in Table.~\ref{Tab:PreformsFeat},
we can find the external quantum efficiency, $\eta_{ext}$, of the samples, which are summarized in Table.~\ref{Tab:Fittings}. Note that the blue lines in Fig.~\ref{Fig:cooleff-All}
are the results of the one-parameter fitting--we could have used the fitting procedure to determine the values of $\alpha_b$ as well. However, the lack of experimental data for
$\eta_c$ at wavelengths above 1070\,nm results in large uncertainties in $\alpha_b$; therefore, we have chosen to use the directly-measured values values in Table.~\ref{Tab:PreformsFeat},
which appear to conform well to our measurements.  

\begin{table}[htp]
   \caption{Results of the fitting procedures related to the LITMoS tests presented in Fig.~\ref{Fig:cooleff-All} with Eq.~\ref{eq:etac}; 
            and the temporal evolution curves of the temperature
            exposed to the high-power Nd:YLF laser presented in Fig.~\ref{Fig:PowCoolTF-All} with Eq.~\ref{Eq:TemporalTemDffeq}.}
\begin{center}
\label{Tab:Fittings}
 \renewcommand{\arraystretch}{1.3}
\begin{tabular}{ | c  | c | c | c | c |}
      \hline
       sample            & $\eta_{ext}$                  & $\Delta T_{max}$ [K]           &  $\tau_c$ [s]    & $\eta_c$ [\%] (1053\,nm)\\
      \hline
       A                 & \hspace*{-2pt}0.993$\pm$0.003 &   0.6                          & 599             & \hspace*{4pt}2.2 \\
      \hline
       B                 & \hspace*{-2pt}0.990$\pm$0.003 &   0.7                          & 754             & \hspace*{4pt}2.7 \\
      \hline
       C                 & \hspace*{-2pt}0.984$\pm$0.003 &   0.56                         & 915             & \hspace*{4pt}2.1 \\ \hline
\end{tabular}
\end{center}
\end{table}

\begin{figure}[ht]
    \includegraphics[width=3.4 in]{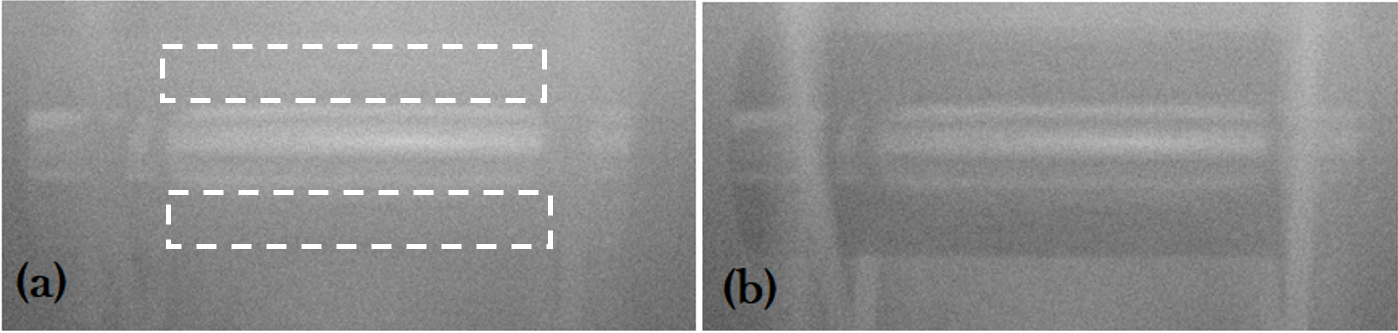}
\caption{{\bf Thermal camera images before and after cooling.} 
         (a) shows the thermal image of samples A before it is exposed to the laser light; 
         (b) shows its thermal image after being cooled by the Nd:YLF laser.
         The regions that are used in obtaining the temperature changes for the LITMoS test are marked by the dashed lines in the left column.
         Thermal images for samples B and C are available in Supplementary Information.}
\label{Fig:PowCool-A}
\end{figure}

The results of the LITMoS tests prove laser cooling in all the Yb-doped silica glass preforms. However, because in the LITMoS test setup, the maximum power of our Ti:Sapphire laser 
in the cooling wavelength range is less than 900\,mW, the signal to noise ratio, as can be seen 
from the error-bars in Fig.~\ref{Fig:cooleff-All}, is large. 
Therefore, to enhance and further confirm the laser cooling of our samples, we pumped the preforms
with a 10.4\,W Nd:YLF laser, the wavelength of which at 1053\,nm resides in the cooling spectral range of the samples (see Fig.~\ref{Fig:cooleff-All}). 
Similar to the LITMoS test, the samples were double-pass pumped by the Nd:YLF laser inside the vacuum chamber and the changes of the temperature were recorded by the thermal camera
as a function of the exposure time. Figure~\ref{Fig:PowCool-A} shows the thermal images of sample A  (a) before and (b) after the exposure to the laser light. 
Subfigure (b) was taken after the laser was turned on and the sample temperature was stabilized ($\sim$40 minutes). Note that the heat extraction occurs 
only in the core of each sample, but the entire sample cools almost uniformly in less than a minute. The cooling is easily recognizable by unaided human eye when the thermal camera 
image become darker after the exposure to the Nd:YLF laser. The bright regions in the thermal image of the sample in Fig.~\ref{Fig:PowCool-A} can be misleading; the reason for these bright regions 
is that silica glass is not transparent in the thermal window and the bright regions on the sample originate from reflections of the thermal radiation from the side-walls of the chamber 
onto our sample's cylindrical surface and eventually into the thermal camera.

\begin{figure}[ht]
\centering
\includegraphics[width=3.4 in]{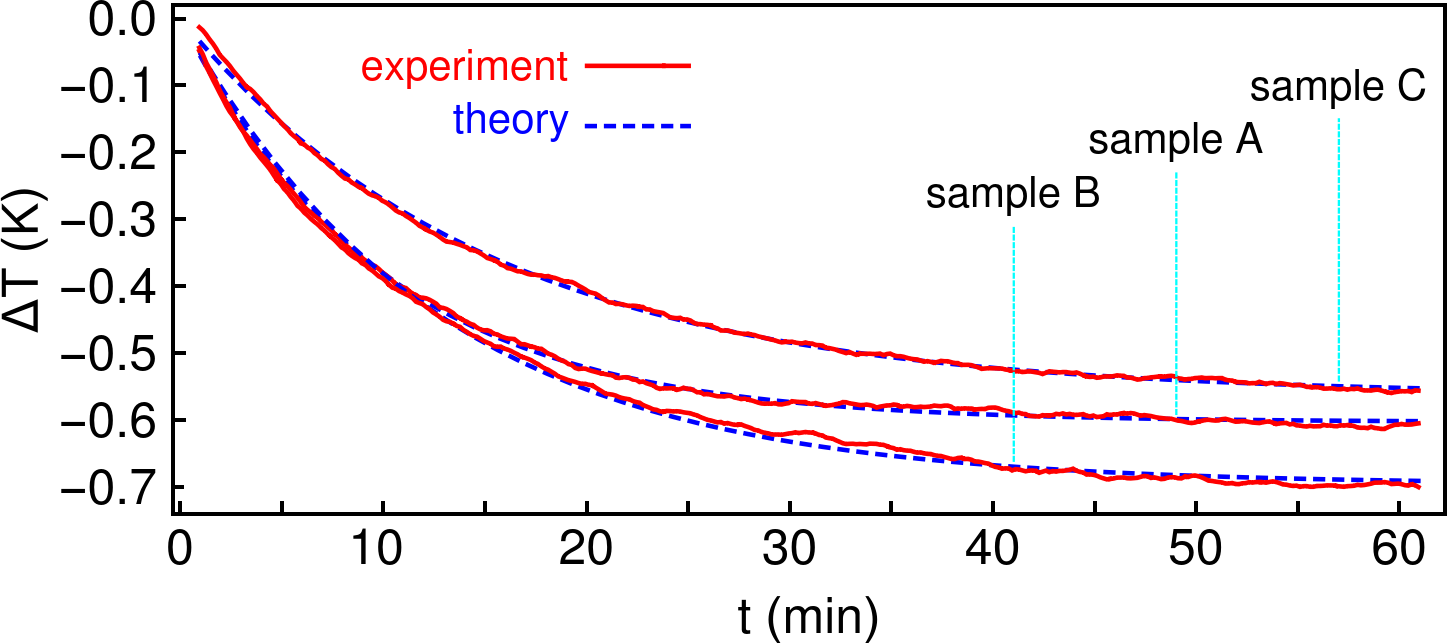}
\caption{{\bf Temporal cooling behavior of the samples.} 
The temperature changes of the samples A, B, and C as a function of time, respectively, when exposed to the Nd:YLF laser. 
The red lines represent the experimental results and the blue lines represent 
the fitting of the exponential function in Eq.~\ref{Eq:TemporalTemDffeq} to the experimental data. 
The obtained $\Delta T_{max}$ and $\eta_{ext}$ parameters from fitting are presented in Table.~\ref{Tab:Fittings}.
}
\label{Fig:PowCoolTF-All}
\end{figure}

Figure~\ref{Fig:PowCoolTF-All} shows the evolution of the temperature of the samples over time while being exposed to the 10.4\,W Nd:YLF laser. 
In each case, the temperature drop can be fitted to the exponential function in Eq.~\ref{Eq:TemporalTemDffeq}:
\begin{align}
\label{Eq:TemporalTemDffeq-Sol}
\Delta T(t) &=\Delta T_{max}(e^{-t/\tau_c}-1),
\end{align}
where we use the following definitions:
\begin{align}
\label{Eq:DTmax}
\Delta T_{max}=\eta_{c}\dfrac{P_{abs}}{4\epsilon \sigma T_{0}^3 A}, \quad\quad \tau_c=\dfrac{\rho V c_{v}}{4\epsilon \sigma T_{0}^3 A}.
\end{align}
$P_{abs}$ is the absorbed power, $\epsilon=0.85$ is the emissivity of the implemented Yb-doped silica glass fiber preforms, 
$\sigma=5.67\times 10^{-8}$~W.m$^{-2}$.K$^{-4}$ is the Stefan-Boltzmann constant,
$T_{0}$ is the ambient temperature, 
$l$ is the sample length, 
$A$ is the surface area of the sample,
$V$ is the volume of the sample, 
$\rho=2.2 \times 10^{3}$~kg.m$^{-3}$ is the silica glass mass density, 
and $c_{v}=741~$J.kg$^{-1}$.K$^{-1}$ is the specific heat of the silica glass.~\cite{yoder,karimi2018theoretical}. 

Equations~\ref{Eq:TemporalTemDffeq-Sol} and~\ref{Eq:DTmax} can be derived by noting that in the vacuum chamber, the convective and conductive heat transfers are negligible; 
therefore, the temporal behavior of the temperature obeys the following differential equation~\cite{Seletskiy:10} (see Supplementary Information):
\begin{align}
\label{Eq:TemporalTemDffeq}
\rho V c_{v}\dfrac{d\Delta T}{dt}\approx -\eta_{c}P_{abs}-4\epsilon \sigma A T_{0}^3 \Delta T,
\end{align}
where the absorbed power in the double-pass experiment is given by
\begin{align}
\label{Eq:Pabsorbed}
P_{abs}=P_{in}\mathcal{T}\big(1-e^{-\alpha_r(\lambda_{p})l}\big)\big(1+\mathcal{T}^{2}R_{m}~e^{-\alpha_r(\lambda_{p})l}\big). 
\end{align}
$\Delta T=T_{s}-T_{0}$, where $T_{s}$ is the sample temperature. $\alpha_r(\lambda_{p})$ is the resonant absorption coefficient of the pump laser.
We also have $\mathcal{T}=T_{w}T_{l}T_{g}$, where $T_{w}=0.92$ is the transmission of the vacuum chamber windows, $T_{l}=0.998$ is the transmission of the lenses, 
$T_{g}=0.96$ is the transmission of the preforms' facets, and $R_{m}=0.998$ is the reflection of the mirror. Note that the absorption coefficients of 
samples~A, B, and C were measured to be $\alpha_r(\lambda_{p})$\,=\,0.43\,m$^{-1}$, 0.52\,m$^{-1}$, and 0.50\,m$^{-1}$, respectively.
The exponential form presented in Eq.~\ref{Eq:TemporalTemDffeq-Sol} is a direct solution to Eq.~\ref{Eq:TemporalTemDffeq};
by fitting Eq.~\ref{Eq:TemporalTemDffeq-Sol} to the measurements in Fig.~\ref{Fig:PowCoolTF-All}, the values of the two fitting parameters
for each sample, i.e., $\Delta T_{max}$ and $\eta_{ext}$ are extracted and reported in Table.~\ref{Tab:Fittings}. 
We note that the slope of the $\Delta T(t)$ curve at $t=0$ in Eq.~\ref{Eq:TemporalTemDffeq-Sol} gives us the value of the cooling efficiency at
1053\,nm wavelength, i.e., $\eta_c=-(\rho l A c_{v}/P_{abs})\partial_t\Delta T|_{t=0}$. For each sample, we also calculate the value of $\eta_c$ 
using the two fitted coefficients, and present the results in Table.~\ref{Tab:Fittings}; these values all agree well, within the error-bar, 
with the plots of $\eta_c$ in Fig.~\ref{Fig:cooleff-All} obtained using the LITMoS tests. Video clips of the cooling evolution of the samples
are available in the Supplementary Information.

In conclusion, we have demonstrated laser cooling in three separate bulk samples of Yb-doped silica glass optical fiber preform. Each sample has a different Yb ion concentration
and each is codoped with one or more of Al, P, F, and Ce elements. We performed a LITMoS test on each sample and extracted its cooling efficiency and showed that
each sample is cooled over a certain laser pump wavelength range. Separately, we exposed each sample to a high-power Nd:YLF laser at 1053\,nm wavelength and monitored 
the temporal evolution of its temperature. The independently extracted cooling efficiencies all agree with those from the LITMoS tests, indicating a maximum cooling
of the three samples by 0.6\,K, 0.7\,K, and 0.56\,K, respectively, at 1053\,nm laser pump wavelength. Because of the geometry of the samples, the temperature variation within 
each sample is negligible; therefore, the reported temperature drop is nearly uniform in the entire volume of each sample~\cite{Mobini:19}. 
The experiments also allowed us to extract the parasitic background absorption
and external quantum efficiency of each sample. We emphasize that this is the first reported measurement of the external quantum efficiency of Yb-doped silica glass,
the determination of which is critical to laser cooling experiments. 

\newpage
\section*{Methods}
\noindent{\bf Fabrication of silica glass preforms.}
The high-purity glasses were fabricated by modified chemical vapor deposition (MCVD) technique. Yb-doping was performed by either the all-solution doping technique (sample A)~\cite{Kuhn:18} 
or gas-phase doping technique (samples B and C)~\cite{S.Kuhn-SPIE-2019}. As stated above, Al codoping was used to ensure a good solubility of the Yb, employing an Al/Yb ratio of greater than 7:1. 
Codoping with P or Ce, as well as the gas-phase doping technique, reduced the photodarkening loss of the material (samples B and C). 
For sample C, a Ce/Yb ratio of about 0.3 was used, which is known to reduce photodarkening substantially~\cite{Jetschke:16}. 

These glasses were developed for single-mode, high-power fiber lasers, so controlling the core-cladding refractive index step was essential; for this reason, codoping with phosphorus~\cite{Kuhn:18}  
or fluorine was used to decrease the refractive index of the material. Fiber lasers using these types of glasses have been used to achieve CW output powers of greater than 4\,kW, 
while maintaining good beam quality~\cite{Beier:18,Beier:17}. Output powers like these can only be accomplished with, among others, high-purity core materials with low background absorption. 
The background losses of these glasses are listed in Table.~\ref{Tab:PreformsFeat} and are less than or equal to 10\,dB/km measured at 1200\,nm wavelength. 
At this wavelength, the main contributions to loss are from Fe$^{2+}$ impurities and Rayleigh scattering. Assuming that Fe$^{2+}$ impurity is the only loss channel, 
its concentration can be estimated to be around 15\,ppb~\cite{schultz1974}; if scattering losses are also considered, the Fe$^{2+}$ concentration would be even lower. 
The OH-induced quenching is negligible because the OH concentration in these glasses is very low (see Table.~\ref{Tab:PreformsFeat}); similar investigations on Yb-doped aluminosilicate 
glasses support this~\cite{Kuhn:15}.
\newpage
\vspace*{3pt}
\section*{Acknowledgements}
The authors would like to acknowledge R. I. Epstein, M. P. Hehlen, and S. D. Melgaard for helpful discussions.
This material is based upon work supported by the Air Force Office of Scientific Research under award number FA9550-16-1-0362 titled Multidisciplinary Approaches to Radiation Balanced Lasers (MARBLE).
\section*{Contributions}
E.M. and A.M. wrote the manuscript and all authors contributed to its final editing. E.M., S.R., and M.P. conducted all the experiments and analyzed the data;
A.A. assisted with the Nd:YLF laser setup. 
S.K., S.H., C.H., J.N., N.H., T.S., and R.E. are responsible for the production and characterization of the silica glass preforms and A.T. supervised their work.
A.M. and M.S.B. led and supervised the laser cooling aspects of the work and participated in the data analysis.
\newpage

\renewcommand{\thepage}{S\arabic{page}} 
\renewcommand{\thesection}{S\arabic{section}}  
\renewcommand{\thetable}{S\arabic{table}}  
\renewcommand{\thefigure}{S\arabic{figure}}
\renewcommand{\theequation}{S\arabic{equation}}


\setcounter{page}{0}
\setcounter{section}{0}
\setcounter{table}{0}
\setcounter{figure}{0}
\setcounter{equation}{0}

{\begin{center}\bf Supplementary Information\end{center}}
\section{Derivation of the cooling efficiency formula}
The cooling efficiency, $\eta_c$, is defined as the net power density (per unit volume) extracted from the material ($p_{\rm net}$) 
per unit power density absorbed or scattered ($p_{\rm abs}$): $\eta_c=p_{\rm net}/p_{\rm abs}$. The net cooling power density can be written as 
$p_{\rm net}=p_{\rm asf}-p_{\rm abs}$, where $p_{\rm asf}$ is the fraction of the  power density that escapes as anti-Stokes fluorescence (ASF) emission out of the material. 
The absorbed power density is given by $p_{\rm abs}=(\alpha_r+\alpha_b)I_P$, where $I_P$ is the pump intensity, $\alpha_r$ is the resonant absorption coefficient of the pump laser 
and, $\alpha_b$ represents the parasitic background absorption. The ASF emission power escaping from the material can be described by $\eta_e N_2 W_r (h\nu_f)$, 
where $\nu_f$ is the mean florescence frequency ($\nu_{f}=c \lambda_{f}^{-1}$, $c$ is the light speed), $N_2$ is the ion density of the excited upper level in the quasi two-level system 
and, $W_r$ ($W_{nr}$) is the radiative (non-radiative) decay rate of the excited state of the doped ions. $\eta_e$ is the extraction 
efficiency and $1-\eta_e$ is the fraction of photons which are trapped inside the host. The rate equation of the energy upper level can be expressed as
\begin{equation}
\dfrac{dN_2}{dt}=\dfrac{\alpha_r I_P}{h\nu_p}-(W_r+W_{nr})N_2+(1-\eta_e)W_rN_2,
\end{equation}
where we have assumed that the trapped florescence finally is reabsorbed by the ions. Under steady-state where $dN_2/dt=0$, the power extracted via ASF emission can 
be written as $p_{\rm asf}=\alpha_r I_P\eta_{ext}(\lambda_p/\lambda_f)$, where the external quantum efficiency is given by $\eta_{ext}=\eta_eW_r/(\eta_eW_r+W_{nr})$. We therefore have
\begin{equation}
p_{\rm net}=\alpha_r I_P\eta_{ext}(\frac{\lambda_p}{\lambda_f})-(\alpha_r+\alpha_b)I_P,
\end{equation}
which leads to 
\begin{equation}
\eta_{c}=\frac{p_{\rm net}}{p_{abs}}=\frac{\lambda_p}{\lambda_f}\eta_{ext}\eta_{abs}(\lambda)-1.
\end{equation}

\section{The non-radiative decay channels in Yb-doped silica glass}
The internal quantum efficiency, $\eta_{q}=W_{r}/(W_{r}+W_{nr})$, is the ratio of the radiative decay to the total decay of an excited state in a medium;
The non-radiative decay channels in a typical Yb-doped silica glass can be divided into a set of decay channels:
\begin{align}
W_{\rm nr}=W_{\rm mp}+W_{\rm OH^{-}}+W_{\rm Yb}+\sum_{\rm i} W_{i}.
\label{Eq:total-nonradiative}
\end{align}
The partial non-radiative decay channels are as follows: 
$W_{\rm mp}$ represents the multiphonon decay of the Yb excited state, 
$W_{\rm OH^{-}}$ accounts for non-radiative decay of the Yb excited state via the high-energy vibrational modes of ${\rm OH^{-}}$ impurities, 
$W_{\rm Yb}$ accounts for non-radiative decay in Yb ion clusters, 
and $W_{i}$ represents the non-radiative decay due to interactions of the excited state with various 
transition-metal and rare-earth ion impurities, respectively.

\begin{figure}[h!]
  \includegraphics[width=\linewidth]{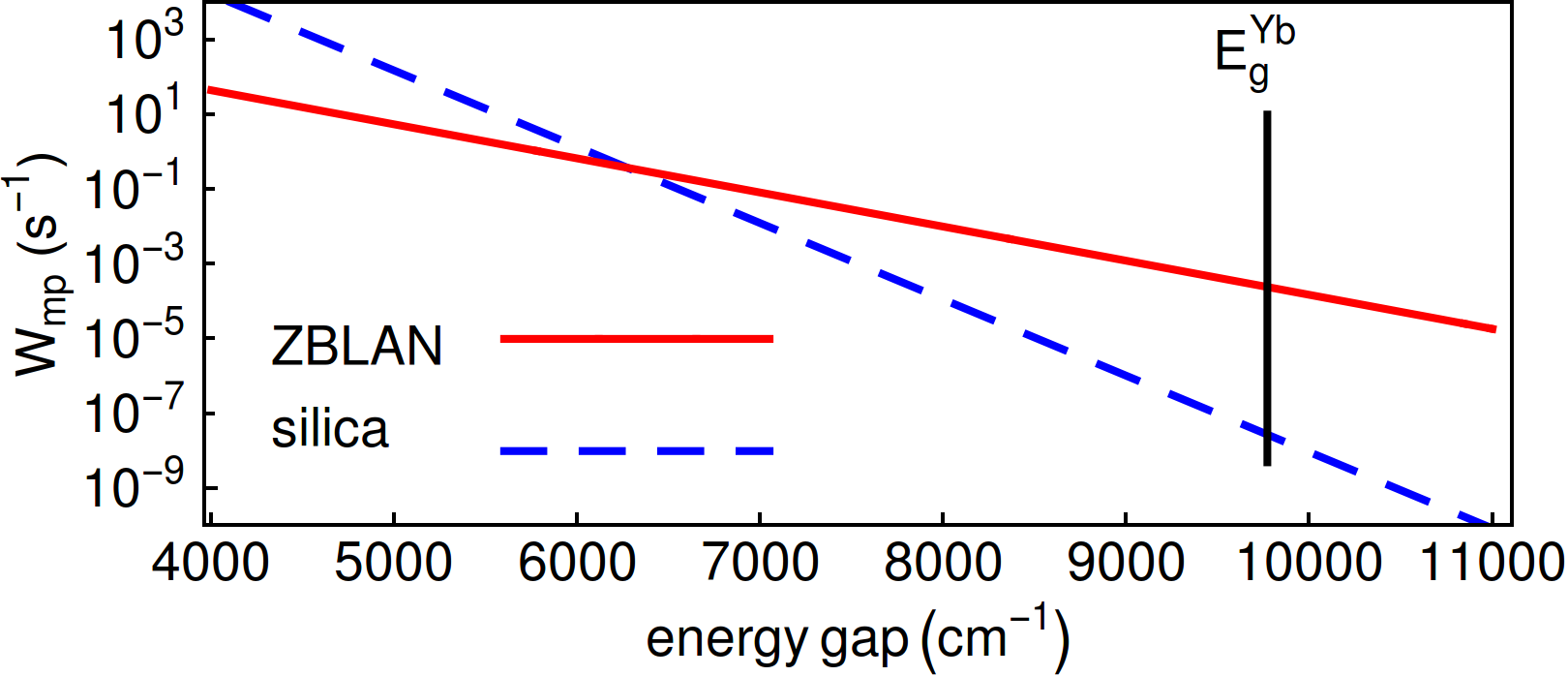}
  \caption{{Non-radiative decay rate of Yb in silica versus ZBLAN glass.} Multi-phonon non-radiative decay rate ($W_{\rm mp}$) of Yb-doped ZBLAN and silica glasses versus energy gap ($E_{g}$)
calculated from Eq.~\ref{Eq:multi-phonon decay} and the parameters listed in Table~\ref{silicazblan}.}
  \label{fig:omeganr}
\end{figure}

 The multiphonon relaxation that originates from the coupling of the excited state with the vibrational 
wavefunctions of the ground state can be described by energy-gap law~\cite{van1983nonradiative,faure2007improvement,hehlen2007model,hoyt2003advances}: 
\begin{align}
\label{Eq:multi-phonon decay}
W_{\rm mp}=W_{0}~e^{-\alpha_{h} ( E_{g}-2E_{p})},
\end{align}
where $E_{p}$ is the maximum phonon energy of the host material, and $E_{g}$ is the energy gap of the dopant ion (Yb).
$W_{0}$ and $\alpha_{h}$ are phenomenological parameters, whose values strongly depend on 
 the host-material~\cite{van1983nonradiative,faure2007improvement,hehlen2007model,hoyt2003advances}. 
Figure~\ref{fig:omeganr} shows the multiphonon non-radiative decay rates of silica and ZBLAN glasses versus the 
energy gaps of the doped ions at $T\,=\,300$\,K, using the parameters shown in Table~\ref{silicazblan}.
\begin{table}[htp]
   \caption{Parameters related to Eq.~\ref{Eq:multi-phonon decay} and Fig.~\ref{fig:omeganr} for silica and ZBLAN glasses~\cite{faure2007improvement,hehlen2007model,hoyt2003advances}.}
\begin{center}
\label{silicazblan}
 \renewcommand{\arraystretch}{1.3}
\begin{tabular}{ | p{10mm} | p{20mm} | p{20mm} | p{15mm} |}
      \hline
      Host & $W_{0}\,({\rm s}^{-1})$ & $\alpha_h\,({\rm cm})$ & $E_p\,({\rm cm}^{-1})$ \\ 
      \hline
       silica & $7.8 \times 10^{7}$   & $4.7 \times 10^{-3}$ & $1.10 \times 10^{3}$ \\
      \hline
      {\small ZBLAN} & $1.7 \times 10^{4}$ & $2.1 \times 10^{-3}$ & $0.58 \times 10^{3}$ \\ \hline
\end{tabular}
\end{center}
\end{table}
As it is obvious from Fig~\ref{fig:omeganr}, the multiphonon relaxation decay rate for excited-state Yb is much smaller in silica glass than in ZBLAN glass.

Auzel et. al. in Ref.~\cite{auzel2003radiation} have shown that the total effect of the last three terms in Eq.~\ref{Eq:total-nonradiative} 
can be described by a phenomenological equation based on a limited diffusion process, 
modeled as a non-radiative dipole-dipole interaction between the ions and impurities~\cite{auzel2003radiation,boulon2008so}:
\begin{align}
\eta_{q}(N)=\dfrac{1}{1+\dfrac{9}{2\pi}\left(\dfrac{N}{N_{c}}\right)^q},
\label{eq:quenchingeq}
\end{align}
where for the glasses, e.g. silica, $q\approx 2$, $N$ is the ion density, and $N_{c}$ is the quenching concentration density~\cite{auzel2003radiation,boulon2008so}.

Equation~\ref{eq:quenchingeq} shows that as long as $N\ll N_{c}$, the internal quantum efficiency approaches unity ($\eta_{q}\approx 1$). All the Yb-doped silica glass
preforms implemented in the study were fabricated in such a way as to satisfy $N \ll N_{c}$, and guarantee a near-unity internal quantum efficiency.
\section{Mean fluorescence wavelength}
\begin{figure}[h!]
  \includegraphics[width=\linewidth]{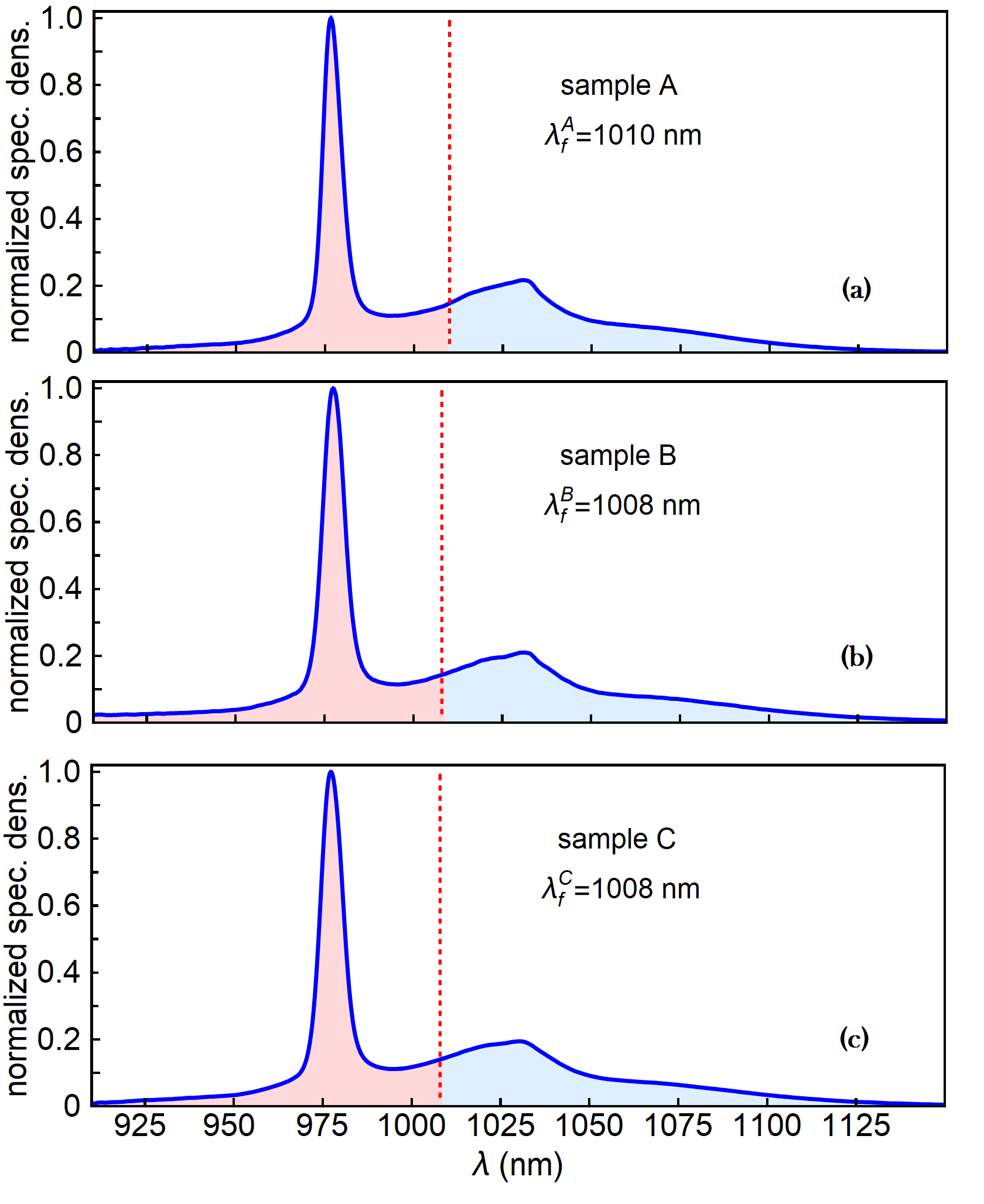}
  \caption{{\bf Normalized spectral density of the samples.} The vertical red dashed line represents the mean fluorescence wavelength ($\lambda_{f}$). 
          (a) The normalized spectral density for sample A versus wavelength with $\lambda_{f}^{A}$=1010\,nm, 
          (b) for sample B with $\lambda_{f}^{B}$=1008\,nm, and 
          (c) for sample C with $\lambda_{f}^{C}$=1008\,nm.}
  \label{fig:meanfwavelength}
\end{figure}

The mean fluorescence wavelength ($\lambda_{f}$) is the average wavelength associated with the average energy of an emitting photon.
 If we assume that $\phi(\nu)$ is the photon flux density, then the average energy of a photon that emits via ASF ($\bar{E}$) takes the following form:
\begin{equation}
\bar{E}=h \nu_{f}=h\frac{\int{\phi(\nu)\nu d\nu}}{\int{\phi(\nu)d\nu}}.
\label{Eq:Meanffreq}
\end{equation}
Considering that $d\nu=-cd\lambda/\lambda^{2}$ and $(hc/\lambda)\,S(\lambda)=h\,\nu\,\phi(\nu)$, where $S(\lambda)$ is the spectral density, the mean fluorescence wavelength can be obtained from:
\begin{align}
\label{Eq:MeanFlWave}
\lambda_{f}=\frac{\int{S(\lambda)\,\lambda\, d\lambda}}{\int{S(\lambda)\,d\lambda}}.
\end{align}
As mentioned earlier, by pumping the silica preforms at $\lambda_{p}$=1030\,nm, the spectral densities of the samples, $S(\lambda)$, were captured by an optical spectrum analyzer (Yokogawa-AQ6319). 
We used Eq.~\ref{Eq:MeanFlWave} to calculate the mean fluorescence wavelength of each sample.
Figure~\ref{fig:meanfwavelength} shows the normalized spectral densities of the three implemented Yb-doped silica samples. 
The mean fluorescence wavelength of each sample is also shown as a legend in each subfigure of Fig.~\ref{fig:meanfwavelength}.
\section{Temperature dynamic of cooling sample under high vacuum}
Four different heat sources can contribute to the temperature changes of the cooling sample. The contribution of each heat-load on the sample can be described by
\begin{align}
\label{Eq:drTmpEv}
&C_{v}\frac{dT}{dt}=-\eta_{c}P_{abs}+\frac{\epsilon A \sigma}{1+\chi}(T_{0}^4-T^4)\\
&+A \kappa_{h} (T_{0}-T)+\frac{N\kappa_{l}A_{l}}{d_{l}}(T_{0}-T)\nonumber,  
\end{align} 
where the first term represents the heat extraction from ASF cooling, the second term represents 
the radiative heat exchange between the cooling sample and the chamber, and the third and forth 
terms represent the convective and conductive heat-loads on the cooling sample, respectively. 
$\eta_{c}$ is the cooling efficiency, $P_{abs}$ is the absorbed power, $\sigma$ is the Stephan-Boltzmann coefficient, 
$T_{0}$ is the ambient temperature, $T$ is the sample temperature ,$\chi=(1-\epsilon_{c})\epsilon A/\epsilon_{c}A_{c}$, 
$\epsilon_{c}$ is the chamber emissivity, $A_{c}$ is the chamber surface area, $N$ is the number of contacting points, 
$A_{l}$ is the area of contacting point, $d_l$ is the length of the contacting point, $\kappa_{l}$ is the thermal conductivity of the sample holder, 
$\kappa_{h}$ is the convective heat transfer coefficient of the chamber and, 
$C_{v}=c_{v}\rho V$, where $V$ is the sample volume and $c_{v}$ is the heat specific coefficient.   

Under high vacuum, the convective heat transfer coefficient becomes negligible~\cite{Seletskiy:10}; therefore, one can ignore the contribution of the convective 
heat source in Eq.~\ref{Eq:drTmpEv}. Similarly, the conductive heat-load from the set of silica fiber holders that are used to support the sample 
are quite low (small $N$ and $A_{l}$). In fact, it is typical for a laser cooling experiment that special care is taken to ensure that the product of 
$\kappa_{l} N A_{l}$ is small such that the contribution of the conductive part becomes negligible~\cite{Seletskiy:10}.

Considering the fact that the surface area of the chamber ($A_{c}$) is much larger than that of cooling sample ($A$), we can assume that $\chi\ll 1$; 
therefore, Eq.~\ref{Eq:drTmpEv} reduces to:
\begin{equation}
C_{v}\frac{dT}{dt}\approx-\eta_{c}P_{abs}+\epsilon A \sigma(T_{0}^4-T^4).
\label{Eq:drTmpEv-1}
\end{equation} 
Assuming that the laser cooling experiment is run in a regime where $T\approx T_0$, which is the case in our experiment,
we will have $(T_{0}^4-T^4)\approx 4 T_{0}^3(T_{0}-T)$; hence, Eq.~\ref{Eq:drTmpEv-1} takes the following from:
\begin{equation}
C_{v}\frac{dT}{dt}\approx-\eta_{c}P_{abs}-4\epsilon A \sigma T_{0}^3 (T-T_{0}).
\label{Eq:drTmpEv-2}
\end{equation} 
\section{LITMoS Test}
In steady state ($dT/dt=0$), Eq.~\ref{Eq:drTmpEv-2} results in a relationship between the cooling efficiency, the absorbed power, and temperature changes of the sample:
\begin{equation}
\eta_{c}(\lambda_p)=\frac{\Delta T(\lambda_p)}{P_{abs}(\lambda_p)C_{rad}},
\label{Eq:LITMoStest}
\end{equation} 
where $C_{rad}\approx 4\epsilon A \sigma T_{0}^3 $.

In the LITMoS test, once the sample temperature is stabilized, the thermal image of the sample at each pump wavelength ($\lambda_{p}$) is captured by a thermal camera--the thermal camera used in the study is Thermal Eye Nanocore 640. Knowing that $\eta_{c}(\lambda_{p})\propto \Delta T(\lambda_{p})/P_{abs}(\lambda_{p})$, after normalizing the thermal images of the sample to the absorbed power at each wavelength, using a fitting procedure, one can extract the proportionality constant and the external quantum efficiency ($\eta_{ext}$). Note that at each wavelength, we have used the spectral density ($S(\lambda)$) as a measure for the absorbed power in the sample as $P_{abs}(\lambda_{p})\propto S(\lambda)$.
\section{High-power laser cooling}
\begin{figure}[!h]
    \includegraphics[width=3.4 in]{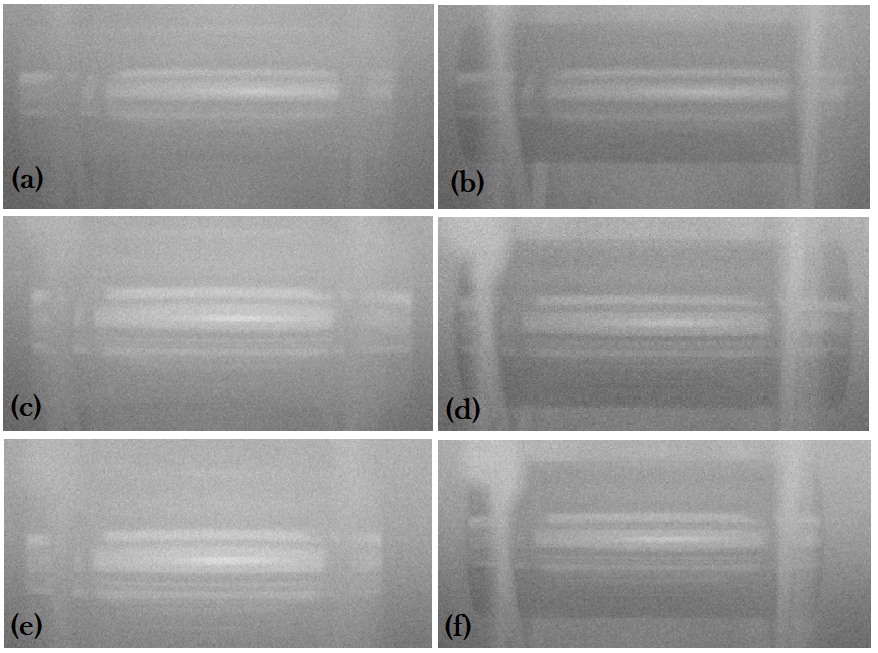}
\caption{{\bf Thermal camera images of the samples.} 
(a), (c) and (e) show the thermal images of the samples A, B and C, respectively, before they are exposed to the laser light. 
         (b), (d) and (f) show the thermal images of the samples A, B and C, respectively, when they are cooled by the Nd:YLF laser.}
\label{Fig:PowCool}
\end{figure}

Figure~\ref{Fig:PowCool} shows the thermal images of the three samples before and after the exposure to the laser light; the three rows starting
from the top correspond to samples A, B, and C, respectively. The images in the left column (a, c and e) were taken before the exposure to the laser and 
the images in the right column (b, d, f) were taken after the laser was turned on and the sample temperatures were stabilized ($\sim$40 minutes). Note that the heat extraction occurs 
only in the core of each sample, but the entire sample cools almost uniformly in less than a minute. The cooling is easily recognizable by unaided human eye when the thermal camera images become darker
after the exposure to the Nd:YLF laser. The bright regions in the thermal images of the samples in Fig.~\ref{Fig:PowCool} can be misleading; the reason for these bright regions 
is that silica glass is not transparent in the thermal window and the bright regions on the samples originate from reflections of the thermal radiation from the side-walls of the chamber 
onto our samples' cylindrical surfaces and eventually into the thermal camera.
\subsection{Videos of high-power laser cooling of samples}
We have enclosed three videos corresponding samples A, B, and C, respectively, that show the temporal evolution of the sample temperatures as captured by the thermal camera
in the high-power laser cooling experiment. As described above, each video shows how the thermal image of the sample gets darker as the sample cools when it is exposed to the
high-power Nd:YLF laser.   

\end{document}